\documentclass[12pt]{article}
\usepackage{graphicx}
\usepackage{epsfig}
\usepackage{setspace}

\title{\bf WMAP 5-year constraints on time variation of $\alpha$ and $m_e$ in a detailed
  recombination scenario}

\author
{  \bf  Claudia G. Sc\'{o}ccola$^{1,2}$\thanks{fellow of
CONICET} , Susana. J. Landau$^{3}$\thanks{member of
the Carrera del Investigador Cient\'{\i}fico y Tecnol\'{o}gico CONICET} , Hector Vucetich$^{1}$ ,
\\ \small$^1$ Facultad de Ciencias Astron\'{o}micas y
Geof\'{\i}sicas. Universidad Nacional de La Plata, \\ \small Paseo del Bosque
S/N 1900 La Plata, Argentina. \\
\small$^2$ Instituto de Astrof{\'\i}sica La Plata - CONICET
\\ \small$^3$ Departamento de F{\'\i}sica, FCEyN, Universidad de
Buenos Aires,\\ \small Ciudad Universitaria - Pab. 1, 1428 Buenos Aires,
Argentina.}

\begin{document}
\maketitle

\begin{abstract}
We study the role of fundamental constants in an updated recombination
scenario. We focus on the time variation of the fine structure
constant $\alpha$, and the electron mass $m_e$ in the early Universe,
and put bounds on these quantities by using data from CMB including
WMAP 5-yr release and the 2dFGRS power spectrum. We analyze how the
constraints are modified when changing the recombination scenario.
\end{abstract}


Time variation of fundamental constants is a prediction of theories
that attempt to unify the four interactions in nature. Many efforts
have been made to put observational and experimental constraints on
such variations. Primordial light elements abundances produced at Big
Bang nucleosynthesis (BBN) and cosmic microwave background radiation
(CMB) are the most powerful tools to study the early universe and in
particular, to put bounds on possible variations of the fundamental
constants between those early times and the present.

Previous analysis of CMB data (earlier than the WMAP five-year
release) including a possible variation of $\alpha$ have been
performed by refs. \cite{Martins02,Rocha03,ichi06,landau08} and including a
possible variation of $m_e$ have been performed by
refs. \cite{ichi06,landau08,YS03}. In March 2008, WMAP team released data collected during
the last five years \cite{Nolta08}. Moreover, new processes relevant during the
recombination epoch have been taken into account. Indeed, in the last
years, helium recombination has been studied in great detail
\cite{Dubrovich05,SH08a,SH08b,SH08c}, revealing the importance of
these considerations on the calculation of the recombination history.
Switzer \& Hirata \cite{SH08a} presented a multi-level calculation for
neutral helium recombination including, among other processes, the
continuum opacity from H {\sc I} photoionization. They found that at
$z<2200$ the increasing H {\sc I} abundance begins to absorb photons
out of the He {\sc I} $2^1p \rightarrow 1^1s$ line, rapidly
accelerating He {\sc I} recombination, which finishes at $z \sim
1800$. Kholupenko et al \cite{KIV07} have considered the effect of the
neutral hydrogen on helium recombination and proposed an approximated
formula to take this effect into account. This has enabled its
implementation on numerical codes such as R{\sc ecfast}, since the
complete calculations done by Switzer \& Hirata take a large amount of
computational time. Another improvement in the recombination scenario is the inclusion of
the semi-forbidden transition $2^3p \rightarrow 1^1s$, the feedback
from spectral distorsions between $2^1p \rightarrow 1^1s$ and $2^3p
\rightarrow 1^1s$ lines, and the radiative line transfer.

The release of new data from WMAP brings the possibility of updating
the constraints on the time variation of fundamental constants.  In
this paper we study the variation of $\alpha$ and $m_e$ in the
improved recombination scenario. It could be argued that $m_e$ is not
a fundamental constant in the same sense as $\alpha$ is and that
constraints on the Higgs vaccum expectation value ($<v>$) are more
relevant than bounds on $m_e$. However, in the recombination scenario
the only consequence of the time variation of $<v>$ is a variation in
$m_e$.


The effect of a possible variation of $\alpha$ and/or $m_e$ in the
recombination scenario and in the CMB temperature and polarization
spectra has been analized previously
\cite{Turner,Hannestad99,KS00,YS03}. Here we focus in the effect of
the variation of $\alpha$ and $m_e$ on the improved recombination
scenario.

The  recombination equations implemented in  R{\sc ecfast}  in the
detailed recombination scenario \cite{wong08} including the
fitting formulae of \cite{KIV07} are:

{\setlength\arraycolsep{1pt}
\begin{equation}
 H(z)(1+z) {dx_{\rm p}\over dz}  =   \Big(x_{\rm e}x_{\rm p} n_{\rm H} 
 \alpha_{\rm H} - \beta_{\rm H} (1-x_{\rm p}) 
   {\rm e}^{-h\nu_{\rm H2s}/kT_{\rm M}}\Big) C_{\rm H}, \\
\end{equation}
\begin{eqnarray}
 H(z)(1+z) {dx_{\rm He II}\over dz} &=&    \left(x_{\rm He II}x_{\rm e} n_{\rm H} \alpha_{\rm HeI}
   - \beta_{\rm HeI} (f_{\rm He}-x_{\rm He II})
   {\rm e}^{-h\nu_{{\rm HeI}, 2^1{\rm s}}/kT_{\rm M}}\right) C_{\rm HeI} \nonumber \\
 &+& \left(x_{\rm He II}x_{\rm e} n_{\rm H} \alpha^{\rm t}_{\rm HeI}
   - \frac{g_{{{\rm HeI}, 2^3{\rm s}}}}{g_{{\rm HeI}, 1^1{\rm s}}} 
   \beta^{\rm t}_{\rm HeI} (f_{\rm He}-x_{\rm He II})
   {\rm e}^{-h\nu_{{\rm HeI},2^3{\rm s}}/kT_{\rm M}}\right)  C_{\rm HeI}^{\rm t}  \ , \nonumber\\
{ } 
\label{eq_triplete}
\end{eqnarray}
\\
where
{\setlength\arraycolsep{1pt}
\begin{eqnarray}
&& C_{\rm H} = {1 + K_{\rm H} \Lambda_{\rm H} n_{\rm H}(1-x_{\rm p})
    \over 1+K_{\rm H} (\Lambda_{\rm H} + \beta_{\rm H})
     n_{\rm H} (1-x_{\rm p}) }, \\
&& C_{\rm HeI} = {1 + K_{\rm HeI} \Lambda_{\rm He} n_{\rm H}
  (f_{\rm He}-x_{\rm He II}){\rm e}^{h\nu_{\rm ps}/kT_{\rm M}}
  \over 1+K_{\rm HeI}
  (\Lambda_{\rm He} + \beta_{\rm HeI}) n_{\rm H} (f_{\rm He}-x_{\rm He II})
  {\rm e}^{h\nu_{\rm ps}/kT_{\rm M}} },  \\
&& C_{\rm HeI}^{\rm t} = {1  \over 1+K^{\rm t}_{\rm HeI}
  \beta^{\rm t}_{\rm HeI} n_{\rm H} (f_{\rm He}-x_{\rm He II})
  {\rm e}^{h\nu^{\rm t}_{\rm ps}/kT_{\rm M}}}.
\end{eqnarray}}

The last term in eq.~(\ref{eq_triplete}) accounts for the
recombination through the triplets by including the semi-forbidden
transition $2^3 {\rm p} \rightarrow 1^1 {\rm s}$.  As remarked in
\cite{wong08}, $\alpha^{\rm t}_{\rm HeI}$ is fitted with the same
functional form used for the $\alpha_{\rm HeI}$ singlets, with
different values for the parameters, so the dependences on the
fundamental constants are the same, being proportional to $\alpha^3 m_e^{-3/2}$.
The two photon transition rates $\Lambda_{\rm H}$ and $\Lambda_{\rm
He}$ depend on the fundamental constants as $\alpha^8m_e$.
The photoionization coefficients $\beta$ are calculated as usual from the
recombination coefficients $\alpha_c$ (with $c$ standing for ${\rm
  H}$, ${\rm HeI}$ and  ${\rm He II}$), so their dependencies
are known (see for example \cite{landau08}).

The $K_{\rm H}$, $K_{\rm HeI}$ and $K^{\rm t}_{\rm HeI}$ quantities
are the cosmological redshifting of the H Ly\,$\alpha$, He\,{\sc i}
$2^1$p--$1^1$s and He\,{\sc i} $2^3$p--$1^1$s transition line photons,
respectively.  In general, $K$ and the Sobolev escape probability
$p_{\rm S}$ are related through the following equations~(taking
He\,{\sc i} as an example): {\setlength\arraycolsep{1pt}
\begin{eqnarray}
&& K_{\rm HeI} = \frac{g_{{\rm HeI}, 1^1{\rm s}}}{g_{{\rm HeI}, 2^1{\rm p}}}
\frac{1}{ n_{{\rm HeI}, 1^1{\rm s}}  
A^{\rm HeI}_{ 2^1{\rm p}-1^1{\rm s}} p_{\rm S}} \quad {\rm and}\\
&& K^{\rm t}_{\rm HeI} = \frac{g_{{\rm HeI}, 1^1{\rm s}}}{g_{{\rm HeI}, 2^3{\rm p}}}
\frac{1}{ n_{{\rm HeI}, 1^1{\rm s}}  
A^{\rm HeI}_{ 2^3{\rm p}-1^1{\rm s}} p_{\rm S}} \ , 
\label{eqKHeI}
\end{eqnarray}}
\\ where $A_{{\rm HeI}, 2^1{\rm p}-1^1{\rm s}}$ and $A_{{\rm HeI},
2^3{\rm p}-1^1{\rm s}}$ are the Einstein $A$ coefficients of the
He\,{\sc I} $2^1$p--$1^1$s and He\,{\sc I} $2^3$p--$1^1$s transitions,
respectively.  To include the effect of the continuum opacity due to
H, based on the approximate formula suggested by ref. \cite{KIV07},
$p_{\rm S}$ is replaced by the new escape probability $p_{esc}=p_s +
p_{\rm con, H}$ with
\begin{equation}
p_{\rm con, H} = \frac{1}{1 + a_{\rm He} \gamma^{b_{\rm He}}}, \\
\end{equation}
and
\begin{equation}
 \gamma = \frac{\frac{g_{{\rm HeI}, 1^1{\rm s}}}{g_{{\rm HeI}, 2^1{\rm p}}}
A^{\rm HeI}_{2^1{\rm p}-1^1{\rm s}} (f_{\rm He} - x_{\rm HeII})c^2}
{8 \pi^{3/2} \sigma_{{\rm H},1{\rm s}}(\nu_{\rm HeI,2^1 p}) 
\nu_{\rm HeI,2^1{\rm p}}^2 \Delta \nu_{\rm D,2^1p} 
(1 - x_{\rm p})}\, 
\end{equation}
where $\sigma_{{\rm H},1s}(\nu_{\rm HeI,2^1p})$ is the H ionization
cross-section at frequency $\nu_{\rm HeI,2^1p}$ and $\Delta \nu_{\rm
D,2^1p} = \nu_{\rm HeI,2^1p} \sqrt{2 k_{\rm B} T_{\rm M}/m_{\rm He}
c^2}$ is the thermal width of the He\,{\sc i} $2^1$p--$1^1$s line. 
The cross-section for photo-ionization from level $n$ is \cite{seaton59}:
\begin{equation}
\sigma_n(Z,h\nu)= \frac{2^6\alpha\pi a_0^2}{3\sqrt{3}} \frac{n}{Z^2}
(1+n^2\epsilon)^{-3} g_{II}(n,\epsilon)
\end{equation}
where $g_{II}(n,\epsilon) \simeq 1$ is the Gaunt-Kramers factor,
and $a_0= \hbar /(m_e c
\alpha)$ is the Bohr radius, so $\sigma_{{\rm H},1s}(\nu_{\rm
HeI,2^1p})$ is proportional to $\alpha^{-1}m_e^{-2}$.

The transition probability rates $A_{{\rm HeI}, 2^1{\rm p}-1^1{\rm
s}}$ and $A_{{\rm HeI}, 2^3{\rm p}-1^1{\rm s}}$ can be expressed as
follows \cite{DM07}:

\begin{equation}
 A^{\rm HeI}_{i-j} = \frac{4 \alpha}{3 c^2} \omega_{ij}^3 \left|\left<\psi_i|r_1 + r_2|\psi_j\right>\right|^2
\label{rate}
\end{equation}
where $\omega_{ij}$ is the frequency of the transition, and $i$($j$)
refers to the initial (final) state of the atom. First we will analize
the dependence of the bra-ket. To first order in perturbation theory,
all wavefunctions can be approximated to the respective wavefunction
of hydrogen. Those can be usually expressed as $\exp(-q r/ a_0)$ where
$a_0$ is the Bohr radius and $q$ is a number. It can be shown that any
integral of the type of Eq.~(\ref{rate}) can be solved with a change of
variable $x=r/a_0$. If the wave functions are properly normalized, the
dependence on the fundamental constants comes from the operator,
namely $r_1 + r_2$. Thus, the dependence of the bra-ket goes as
$a_0$. On the other hand, $\omega_{ij}$ is proportional to the
difference of energy levels and thus its dependence on the fundamental
constants is $\omega_{ij} \simeq m_e \alpha^2$. Consequently, the
dependence of the transition probabilities of ${\rm HeI}$ on $\alpha$ and
$m_e$ is

\begin{equation}
 A^{\rm HeI}_{i-j} \simeq m_e \alpha^5
\end{equation} 

\begin{figure}[!ht]
\begin{center}
\includegraphics[scale=0.3,angle=-90]{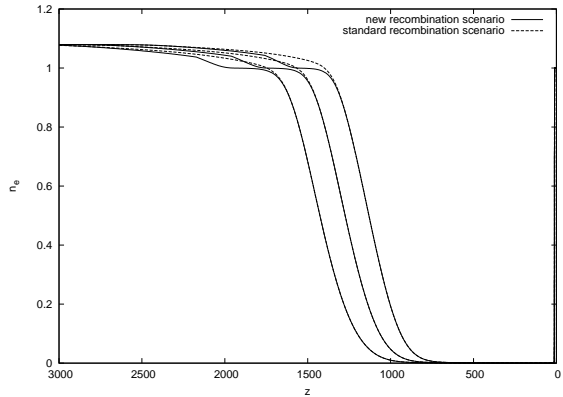}
\end{center}
\caption{Ionization history allowing $\alpha$ to vary with time. From
left to right, the values of $\frac{\alpha}{\alpha_0}$ are 1.05, 1.00,
and 0.95, respectively. The dotted lines correspond to the standard
recombination scenario, and the solid lines correspond to the updated
one.}
\label{ionization_history}
\end{figure}

In Fig.~\ref{ionization_history} we show how a variation in the value
of $\alpha$ at recombination affects the ionization history, moving
the redshift at which recombination occurs to earlier times for larger
values of $\alpha$. The difference between the functions when the two
different recombination scenarios are considered, for a given value of
$\alpha$, is smaller than the difference that arise when varying the
value of $\alpha$. Something similar happens when varying $m_e$.

With regards to the fitting parameters $a_{\rm He}$ and $b_{\rm He}$,
since detailed calculation of their values are not available yet, it
is not possible to determine the effect that a variation of $\alpha$
or $m_e$ would have on these new parameters. Wong et al \cite{wong08}
have shown that they must be known at the $1\%$ level for future
Planck data. In this letter, however, we are dealing with the 5 yr
data from WMAP satellite and this accuracy is not required. To come to
this conclusion, we have calculated the temperature, polarization and
cross correlation CMB spectra, allowing the parameters $a_{\rm He}$
and $b_{\rm He}$ to vary at the 50\% level. We found that for the
temperature and polarization spectra, the variation is always lower
than the observational error (1\% for temperature and almost 40 \% in
polarization). The largest variations occur in the cross correlation
CMB spectra ($C_{\ell}^{TE}$). In this case, we have calculated the
observational errors divided by the value of the $C_{\ell}$'s of all
measured $C_l^{TE}$ and compared them with the relative variation in
the $C_l$'s induced when changing $a_{\rm He}$ and $b_{\rm He}$ by a
$50\%$. In all of the cases the first quantity is several orders of
magnitude greater than the variation of the $C_{\ell}$'s. Therefore,
in order to analyze WMAP5 data, there is no need to modify these
parameters.


To put constraints on the variation of $\alpha$ and $m_e$ during
recombination time in the detailed recombination scenario studied
here, we introduced the dependencies on the fundamental constants
explicitly in the latest version of R{\sc ecfast} code
\cite{recfast}, which solves the recombination equations.  We
performed our statistical analysis by exploring the parameter space
with Monte Carlo Markov chains generated with the publicly available
CosmoMC code of ref.~\cite{LB02} which uses the Boltzmann code CAMB
\cite{LCL00} and R{\sc ecfast} to compute the CMB power spectra. We
modified them in order to include the possible variation of $\alpha$
and $m_e$ at recombination. We ran eight Markov chains and followed
the convergence criterion of ref. \cite{Raftery&Lewis} to stop them
when $R-1< 0.0180$. Results are shown in Table \ref{tablacmb} and
Figure \ref{resulcmb}.

The observational set used for the analysis was data from the WMAP
5-year temperature and temperature-polarization power spectrum
\cite{Nolta08}, and other CMB experiments such as CBI \cite{CBI04},
ACBAR \cite{ACBAR02}, and BOOMERANG \cite{BOOM05_polar,BOOM05_temp},
and the power spectrum of the 2dFGRS \cite{2dF05}. We have considered
a spatially-flat cosmological model with adiabatic density
fluctuations, and the following parameters:
\begin{equation}
P=\left(\Omega_B h^2, \Omega_{CDM} h^2, \Theta, \tau, \frac{\Delta
\alpha}{\alpha_0}, \frac{\Delta m_e}{(m_e)_0}, n_s, A_s\right)\, \, ,
\end{equation}
where $\Omega_{CDM} h^2$ is the dark matter density in units of the
critical density, $\Theta$ gives the ratio of the comoving sound
horizon at decoupling to the angular diameter distance to the
surface of last scattering, $\tau$ is the reionization optical
depth, $n_s$ the scalar spectral index and $A_s$ is the amplitude of
the density fluctuations.

In Fig.~\ref{resulcmb} we show the marginalized posterior
distributions for the cosmological parameters, $\Delta \alpha
/\alpha_0$, and $\Delta m_e/(m_e)_0$, which are the variation in the
values of those fundamental constants between recombination epoch and
the present. The three succesively larger two dimensional contours in
each panel correspond to the 68\%-, 95\%-, and 99\%- probability
levels, respectively. In the diagonal, the one dimensional likelihoods
show the posterior distribution of the parameters.

\begin{figure}[!ht]
\begin{center}
\includegraphics[scale=1.0,angle=0]{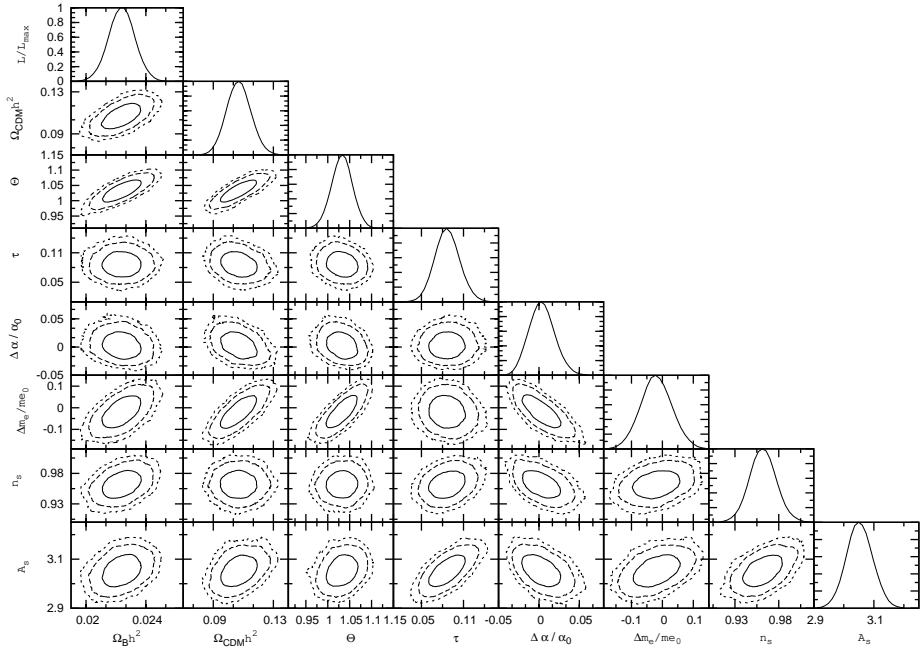}
\end{center}
\caption{Marginalized posterior distributions obtained with CMB data,
  including the WMAP 5-year data release plus 2dFGRS power
  spectrum. The diagonal shows the posterior distributions for
  individual parameters, the other panels shows the 2D contours for
  pairs of parameters, marginalizing over the others.}
\label{resulcmb}
\end{figure}

In Table~\ref{tablacmb} we show the results of our statistical
analysis, and compare them with the ones we have presented in
ref.~\cite{landau08}, which were obtained in the standard
recombination scenario (i.e. the one described in \cite{Seager00},
which we denote PS), and using WMAP3 \cite{wmap3_temp,wmap3_pol} data.
The constraints are tighter in the current analysis, which is an
expectable fact since we are working with more accurate data from
WMAP.  The bounds obtained are consistent with null variation, for
both $\alpha$ and $m_e$, but in the present analysis, the $68 \%$
confidence limits on the variation of both constants have changed. In
the case of $\alpha$, the present limit is more consistent with null
variation than the previous one, while in the case of $m_e$ the single
parameters limits have moved toward lower values. To study the origin
of this difference, we perform another statistical analysis, namely
the analysis of the standard recombination scenario (PS) together with
WMAP5 data and the other CMB data sets and the 2dFGRS power
spectrum. The results are also shown in Table~\ref{tablacmb}. We see
that the changes in the results are due to the new WMAP data set, and
not to the new recombination scenario. In Fig.~\ref{compara} we
compare the probability distribution for $\Delta \alpha/\alpha_0$ and
also for $\Delta m_e/(m_e)_0$, in different scenarios and with
different data sets.

\begin{table}
\renewcommand{\arraystretch}{1.5}
\begin{tabular}{|c|c|c|c|}
\hline
  parameter & wmap5 + NS & wmap5 + PS & wmap3 + PS\\
\hline
 $\Omega_b h^2$ & 0.02241$_{-0.00084}^{+0.00084}$   & 0.02242$_{-0.00085}^{+0.00086}$ & 0.0218$_{-0.0010}^{+0.0010}$ \\
\hline
 $\Omega_{CDM} h^2$ & 0.1070$_{-0.0078}^{+0.0078}$  & 0.1071$_{-0.0080}^{+0.0080}$ & 0.106$_{-0.011}^{+0.011}$ \\
\hline 
$\Theta$ & 1.033$_{-0.023}^{+0.023}$  & 1.03261$_{-0.023}^{+0.024}$ & 1.033$_{-0.029}^{+0.028}$ \\
\hline
$\tau$ & 0.0870$_{-0.0081}^{+0.0073}$   & 0.0863$_{-0.0084}^{+0.0077}$ & 0.090$_{-0.014}^{+0.014}$ \\
\hline
$\Delta \alpha / \alpha_0$ & 0.004$_{-0.015}^{+0.015}$  & 0.003$_{-0.015}^{+0.015}$  & -0.023$_{-0.025}^{+0.025}$\\
\hline 
$\Delta m_e /(m_e)_0$ & -0.0193$_{-0.049}^{+0.049}$  & -0.017$_{-0.051}^{+0.051}$ & 0.036$_{-0.078}^{+0.078}$\\
\hline
$n_s$ & 0.962$_{-0.014}^{+0.014}$  & 0.963$_{-0.015}^{+0.015}$  & 0.970$_{-0.019}^{+0.019}$\\
\hline
$A_s$ & 3.053$_{-0.041}^{+0.042}$ & 3.052$_{-0.043}^{+0.043}$   & 3.054$_{-0.073}^{+0.073}$ \\
\hline
$H_0$ &  70.3$_{-5.8}^{+5.9}$  &   70.3$_{-6.0}^{+6.1}$  &  70.4$_{-  6.8}^{+  6.6}$  \\
\hline
\end{tabular}
\caption{Mean values and 1$\sigma$ errors for the parameters
including $\alpha$ and $m_e$ variations. NS stands for the new
recombination scenario, and PS stands for the previous one.} \label{tablacmb}
\end{table}

\begin{figure}[!ht]
\begin{center}
\includegraphics[scale=0.8,angle=0]{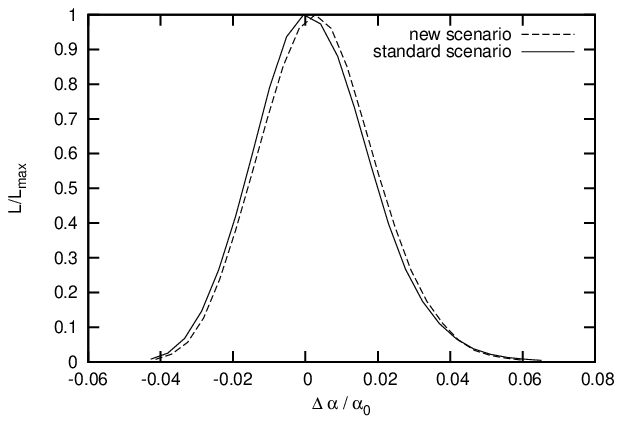}
\includegraphics[scale=0.8,angle=0]{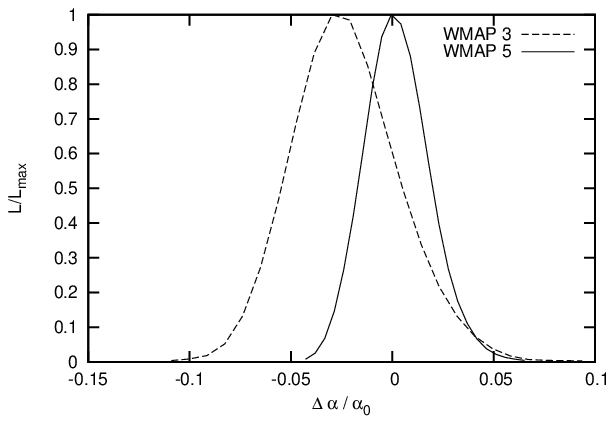}
\includegraphics[scale=0.8,angle=0]{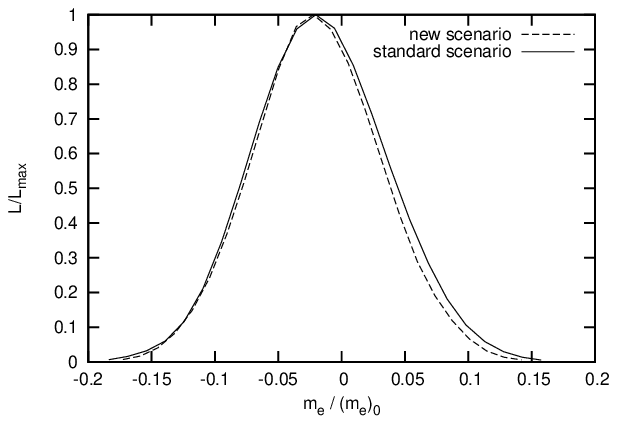}
\includegraphics[scale=0.8,angle=0]{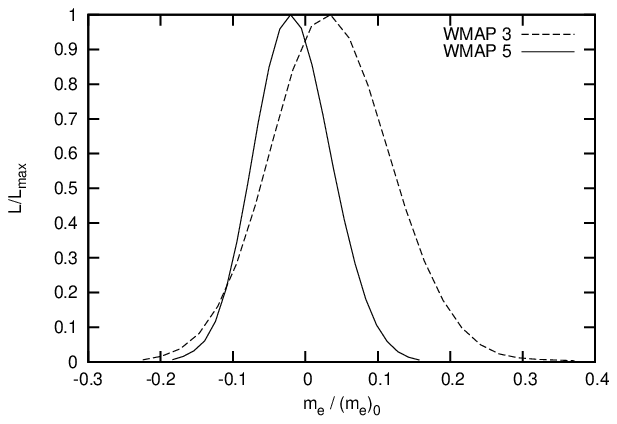}
\end{center}
\caption{One dimensional likelihood for $\frac{\Delta
    \alpha}{\alpha_0}$ (upper row) and $\frac{\Delta
    m_e}{(m_e)_0}$ (lower row). Left: for WMAP5 data and two different
  recombination scenarios.  Right: comparison for the standard
  recombination scenario, between the WMAP3 and WMAP5 data sets.}
\label{compara}
\end{figure}

In Fig.~\ref{resulcmb2} we compare the 95\%- probability contour level
for the parameters, and their one dimensional distributions, for two
different analysis in the standard recombination scenario, namely the
one with WMAP5 data (dashed lines) and the one with WMAP3 data (solid
lines).  The contours are smaller in the former case, which is
expectable since that data set is more accurate. For the fundamental
constants, the contours notably shrink. Moreover, the constraints are
shifted to a region of the parameter space closer to that of null
variation in the case of $\alpha$. On the other hand, limits on the
variation of $m_e$ are shifted to negative values, but still
consistent with null variation. From the one dimensional likelihoods
we see that the the peak of the likelihood has moved for $\Omega_b
h^2$. The obtained results for the cosmological parameters are in
agreement within $1 \sigma$ with the ones obtained by the WMAP
collaboration \cite{wmap5}, without considering variation of
fundamental constants.

\begin{figure}[!ht]
\begin{center}
\includegraphics[scale=1,angle=0]{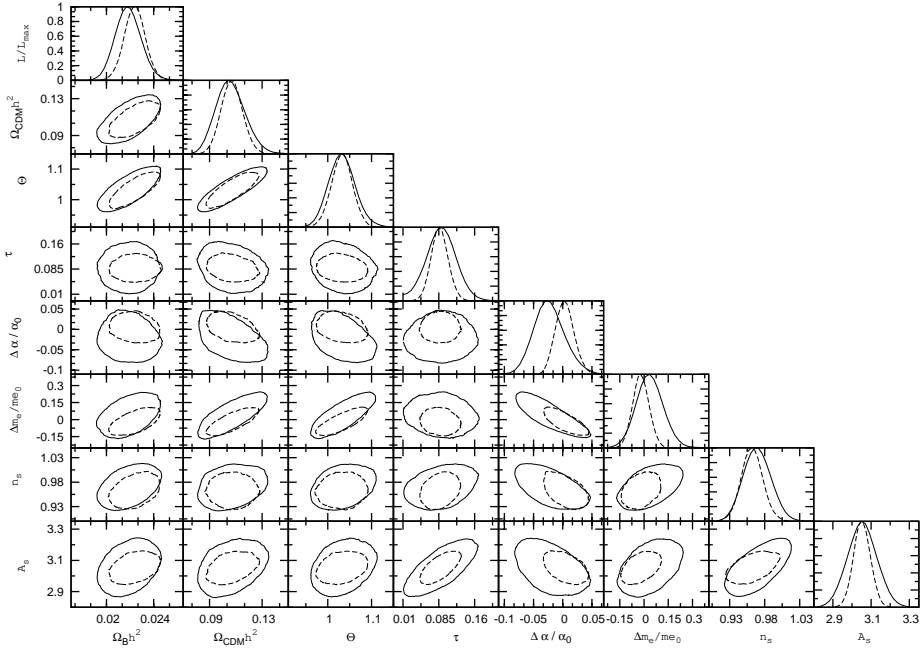}
\end{center}
\caption{Comparison between the 95\%- confidence levels of WMAP3 (solid
  line) with those of WMAP5 (dashed line). In the diagonal, we compare
  the one dimensional likelihoods in these two cases.}
\label{resulcmb2}
\end{figure}

\section*{{\bf Acknowledgments}}

Support for this work was provided by PIP 5284 CONICET.  Numerical
calculations were performed in KANBALAM cluster at Universidad
Nacional Aut\'onoma de M\'exico. The authors would like to thank E.C
Flores, I.R.Leobardo, J.L.Gordillo Ruiz and S. E. Frausto del Rio for
technical support for the KANBALAM facilities.

\end{document}